# Evidence of boron pairs in highly boron laser doped silicon

Léonard Desvignes[1], Francesca Chiodi[1*], Géraldine Hallais[1], Dominique Débarre[1], Giacomo Priante[2], Feng Liao[2], Guilhem Pacot[2], Bernard Sermage[2]

1. Université Paris-Saclay, CNRS, Centre de Nanosciences et de Nanotechnologies, Palaiseau 91120, France
2. Probion Analysis, Parc Technopolis, Bat Alpha, 3 avenue du Canada, 91490 Les Ulis France

* francesca.chiodi@c2n.upsaclay.fr
+ 0033 (0)6 88587005

Abstract

Secondary Ions Mass Spectroscopy and Hall effect measurements were performed on boron doped silicon with concentration between 0.02 at.% and 12 at.%. Ultra-high boron doping was made by saturating the chemisorption sites of a Si wafer with BCl$_3$, followed by nanosecond laser anneal (Gas Immersion Laser Doping). The boron concentration varies thus nearly linearly with the number of process repetitions. However, it is not the case for the hole concentration which tends to saturate at high boron concentration. The difference between boron and hole concentration increases as the square of boron concentration, pointing towards the formation of boron pairs as the dominant contribution to the increase of inactive boron.

## I. Introduction

Highly Boron doped silicon ( > $10^{21}$ cm$^{-3}$) is of great interest as well for its superconducting properties [1,2,3,4,5] as for its use in components [6], for example for source and drain regions in MOSFET transistors. These layers can be made by gas source molecular beam epitaxy, for active concentrations up to $1.3 \times 10^{21}$ cm$^{-3}$ [6], implantation followed by laser annealing [3,7] or Gas Immersion Laser Doping (GILD), up to $6 \times 10^{21}$ cm$^{-3}$ [8,9]. However, at high boron doping level, the hole concentration tends to saturate while the B content keeps growing.

In this paper, we have studied by Secondary Ions Mass Spectroscopy (SIMS) and Hall effect highly boron doped silicon layers with boron concentrations between $10^{19}$ cm$^{-3}$ and $6 \times 10^{21}$ cm$^{-3}$. The layers were realised by GILD, a technique which insures a high versatility in the doping concentration and thickness, and achieves large active concentrations while avoiding potential bias due to implantation defects.

We analyse the B incorporation as a substitutional dopant vs. interstitial defect, and show that for B atomic concentrations as high as 12 at.%, the formation of B dimers in substitutional sites dominates over interstitial B and aggregates.

## II. Sample fabrication and measurement.

GILD doping was performed in a UHV reactor with ultra high residual vacuum (~ $10^{-9}$ mbar) in the reaction chamber. At a frequency of 2 Hz, a puff of boron precursor gas (BCl$_3$) is injected over a (100) oriented silicon surface, saturating the chemisorbtion sites, followed by an excimer laser pulse irradiation (λ = 0.308 µm, 25 ns). The laser doped areas (2mm x 2mm) were subjected to 1-700 sequences of one gas injection followed by one laser pulse (laser fluence 1.3 J/cm$^2$, corresponding to a 175 nm depth melted layer as measured by STEM and SIMS, while the silicon melting threshold is 0.6 J/cm$^2$). Due to the low coherence of the excimer laser, there are no speckles in the beam energy density. On the other hand, appropriate laser optics allows to strongly reduce the spatial inhomogeneity of the transverse laser energy profile to 1% [8]. The melted layer stays liquid during 60 ns as observed by time resolved reflectivity measurements. In the liquid phase, boron diffusion is very fast (~ 10$^{-4}$ cm$^2$/s) as compared to its diffusion in solid Si (~ 10$^{-11}$ cm$^2$/s), so that boron concentration is homogeneous in depth and doping is limited to the melted region. At the end of the laser pulse, the cooling of the liquid phase makes an epitaxy from the substrate toward the surface of the sample. The solid-liquid interface velocity (~ 4 m/s [10]) is extremely fast, allowing large active B concentrations larger than the solubility limit. The segregation coefficient of boron is close to 1 so that the boron profile in the layer is nearly flat with a small increase at the bottom of the layer as can be seen on the SIMS profiles, and with a sharp Si:B/Si interface a few nanometers thick (Fig. 1). The boron increase at the bottom of the layers, observed in SIMS profiles, corresponds to inactive B precipitates at the turning point of the liquid/solid interface, where the velocity is not high enough to overcome the solubility limit. However, such tiny inactive B quantity does not impact significantly the following quantitative study as it accounts for at maximum 1% of the total B.

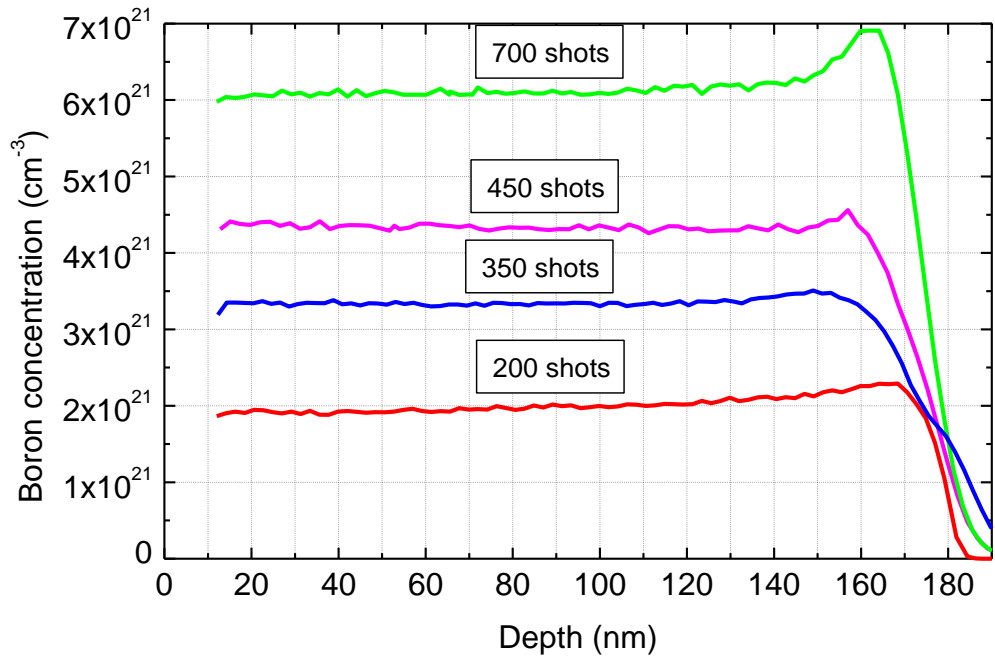

*Figure 1. Boron profile in the doped layer measured by SIMS on spots with 200, 350, 450 and 700 shots. We have suppressed the 15 first nm because of possible artefacts due to the equilibration of the oxidation of silicon by the primary oxygen ions.*

Similar GILD made spots have been extensively analysed by different techniques: resistivity and Hall electrical measurements, scanning transmission electron microscopy (STEM) for structural informations, SIMS and Atom Probe Tomography (APT) to observe the dopant distribution and presence of aggregates. These studies show the reproductiveness of the layers thickness, t=175 nm, and concentration with a 10% precision. Up to concentrations of $2 \times 10^{21}$ cm$^{-3}$ the doping increases linearly by each new laser shot [10] and X-ray measurements show that the layer is strained with the same in plane parameter as the substrate [2,10]. APT [11] and STEM show that there are no boron clusters, that the layer is a random substitutional alloy, that there are no chlorine atoms brought by the dopant gas, as Cl segregation coefficient is near zero so that all Cl atoms are pushed out of the surface. SIMS profiles performed on 4 spots with the same doped layer depth of about 175 nm and 200, 350, 450 and 700 laser shots are shown on Figure 1.

Measurements were done on a 4F Cameca system equipped with a magnetic mass spectrometer. For doping levels above $5 \times 10^{20}$ cm$^{-3}$, the doping value can be suspected due to matrix effect. This effect was studied in detail by C. Dubois et al. [12]. They show that with oxygen primary beam, the variation of the relative ion yield is the same for Boron and Silicon so that the matrix effect is corrected by comparing with the silicon signal and using the Relative Sensibility Factor (RSF). Moreover we analyse only secondary ions at high energy (> 100 eV) because they are less sensitive to the chemical surrounding. Boron concentration is obtained by comparison with a standard from the National Institute of Standards (NIST). Depth calibration is obtained with a mechanical profilometer and we take into account the sputter rate difference between the doped layer and the silicon substrate. The doping level as a function of the number of laser shots is given on Fig. 2 and shows that the variation is linear for small shot numbers, which is expected as we incorporate the same amount of boron atoms due to the saturation of the chemisorption sites on the surface of the silicon sample. Above 500 shots there is a small negative curvature.

The hole concentration was measured by Hall effect on similar spots made on high resistivity n-type silicon substrate. A Hall cross was etched on each spot and the Hall voltage $V_H$ of the layer was measured. The influence of the substrate is negligible due to the n-p barrier between the p-type layer and the n-type substrate. The hole concentration [h] is given by [13,14]:

$$[h] = \gamma \frac{1}{qR_H} \qquad (1)$$

Where q is the electron charge, $R_H$ is the Hall coefficient and γ is the Hall mobility factor which is the ratio between the Hall mobility $\mu_H$ and the conductivity mobility $\mu_c$: γ = $\mu_H/\mu_c$ . This factor is related to the non parabolicity and the warping of the valence band. Taking into account the three higher valence bands, Lin et al. [14] have calculated that γ decreases slightly with the increase of hole concentration and that it is nearly constant above $1 \times 10^{18}$ cm$^{-3}$ and equal to 0.75. The experimental value of γ at high boron doping level depends on the measured mobility [15] : γ = 0.7 from measurements by Thurber et al. [16] and γ = 0.8 from measurements by Irvin [17]. In the following, we take γ = 0.75 which is an average value. On Figure 2, all the boron atoms are electrically active for concentration smaller than $5 \times 10^{20}$ cm$^{-3}$. At higher boron concentrations, the hole concentration has a slower increase than the boron concentration and tends to saturate. The difference between boron and hole concentrations shown on Fig. 2 increases quadratically.

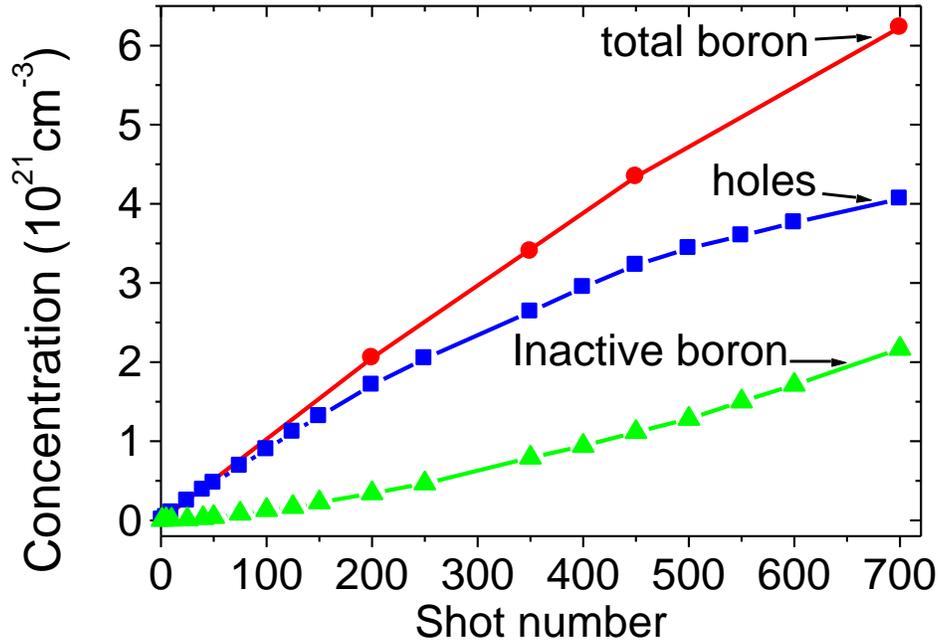

*Figure 2. Variation with the number of shots of the Boron concentration measured by SIMS (red disks), of the hole concentration measured by Hall effect with γ = 0.75 (blue squares) and of the difference between these two concentrations (green triangles).*

This is better seen in Figure 3 which compares the difference between boron and hole concentration to a quadratic variation. The quadratic increase of inactive B atoms suggests the formation of B dimers. These Boron pairs can either be formed by substitutional and interstitial boron ($B_S$-$B_I$) or by two substitutional boron atoms ($B_S$-$B_S$). Since the energy of a $B_S$-$B_I$ pair is larger than that of a $B_S$-$B_S$ pair [18], we will consider only substitutional boron pairs. In conductivity measurements, substitutional boron pairs give only one free hole. Indeed a boron pair links 2 holes: one has nearly the same binding energy as in the case of a single boron atom since it is linked to only one positive charge (46 meV) [19]. This hole is free at room temperature. The second hole is linked to a double charged boron pair and has about a four time larger binding energy (176 meV) [20] without taking into account central cell correction. This hole remains linked at room temperature and is not observed in conductivity and Hall effect, whch explains the difference between bore concentration and hole concentration as due to boron pairs.

In the absence of attraction or repulsion between substitutional boron atoms, the density of boron pairs can be calculated easily. Around each substitutional boron atom, there are 4 places for an eventual other boron atom. The probability to have another boron atom on one of these places is proportional to the concentration of boron atoms [B] divided by the sum of the concentration of silicon [Si] and boron atoms which is equal to $5 \times 10^{22}$ cm$^{-3}$ [13]. Then the concentration of boron pairs [BB] is:

$$[BB] = \frac{4[B]^2}{5.10^{22}} \quad (2)$$

Figure 3 shows that the difference between boron and hole concentrations is close to the concentration of pairs given by equation 2 which proves that the non electrically active boron atoms in the layers are mainly due to substitutional boron pairs and not interstitials or clusters which would have no reason to vary as the square of boron concentration. This agrees with results of APT measurements [11].

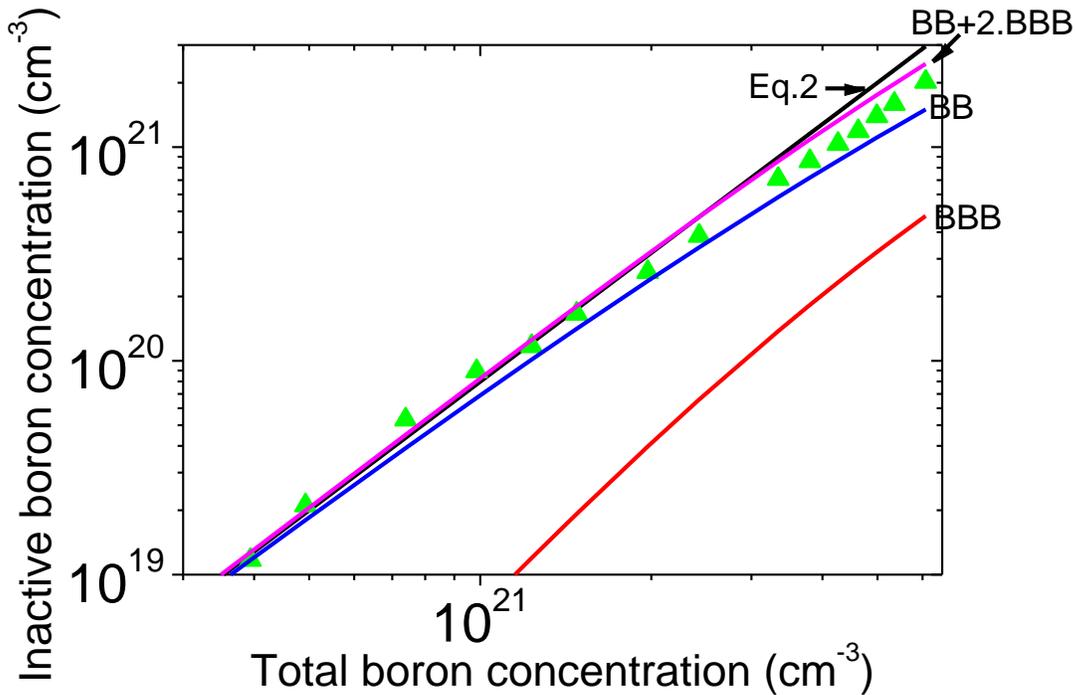

*Figure 3. Difference between boron atoms concentration and hole concentration versus boron concentration (green squares) with γ= 0.75. The black line is the boron pair concentration given by equation 2. The blue line is the concentration of only boron pairs given by equation 4. The red line is the concentration of boron triplets given by equation 7. The magenta line is the sum of the boron pairs and two times the boron triplet concentrations.*

Furthermore, the presence of boron pairs shows that there is no attraction or Coulomb repulsion between boron atoms during the fabrication of the doped layer.
Indeed the melted phase has a metallic behaviour and 70% of the covalent bonds between silicon atoms do not exist [21], so that silicon and boron atoms are positively charged and are immersed in a cloud of about $1.4 \times 10^{23}$ cm$^{-3}$ electrons. Positively charged silicon and boron atoms move rapidly with no electric repulsion because of the large quantity of electrons so that boron ions have perfectly random positions with the same probability to be close to another boron ion or to a silicon ion.

At the end of the laser pulse (25 ns), the temperature decreases and there is an epitaxy of monocrystaline silicon from the bottom interface toward the sample surface with a velocity of about 4 m/s. This fast epitaxy velocity and the high carrier concentration in the crystal result in the same random repartition of boron atoms in the crystal as in the liquid, Substitutional boron pairs exist in the silicon crystal as if there were no repulsion between boron atoms.

Equation 2 is an approximate calculation of boron pair concentration since the first boron atom and the one which is added may already belong to boron pairs so that we have to subtract boron atoms which belong already to boron pairs:

$$[BB] = \frac{4([B]-[BB])^2}{5.10^{22}} \qquad (3)$$

Which gives for the concentration of pairs:

$$[BB] = \frac{4[B]^2}{5.10^{22} + 8[B]} \qquad (4)$$

The value given by equation 4 is shown on figure 3. It is also close to the experimental values. It is smaller than the concentration given by equation 2 which is the sum of pairs, triplets, quadruplets etc… Triplets participate also to the non active boron concentration since each triplet gives only one free hole and two deeply linked holes so that the concentration of free holes is:

$$[h] = [B] - [BB] - 2[BBB] \qquad (5)$$

The concentration of three boron atoms together can be calculated by adding a boron atom to a boron pair with 6 possible places for the third atom:

$$[BBB] = \frac{6([BB]-[BBB])([B]-[BB]-[BBB])}{5.10^{22}} \qquad (6)$$

which gives :

$$[BBB] = \frac{6[BB]([B]-[BB])}{5.10^{22} + 6[B]} \qquad (7)$$

Figure 3 shows the curve of [BB] + 2 [BBB] which is close to the experimental curve of inactive boron but not closer that pairs only. We are convinced on the existence of boron pairs but the existence of boron triplets should not be excluded, as it depends on the precision of the measurements (~ 5%) and on the value of the Hall mobility factor γ.

In conclusion, the inevitable existence of boron pairs and boron triplets limits the concentration of holes obtained by high boron doping in silicon crystals.

## III. Conclusion

The comparison of boron concentration measured by SIMS and hole concentration measured by Hall effect in highly boron doped silicon layers obtained by GILD show that these layers contain boron pairs. The existence of substitutional boron pairs is made possible by the screening of the repulsive electric force due to the high electron density in the melted phase and the high hole density during the crystallisation time of the GILD process. Each boron pair is responsible for only one free hole in the valence band which limits the hole doping in these layers. Hole saturation is thus unavoidable in silicon crystals at high B concentrations as stemming from geometric arguments. The experimental results show that electrically active boron, boron pairs and perhaps boron triplets are sufficient to explain SIMS and Hall measurements for boron concentrations as high as $6 \times 10^{21}$ cm$^{-3}$ i.e. that interstitials and aggregate boron are in much smaller concentration.

## Acknowledgements

FC and DD acknowledge support from the Réseau RENATECH, Plateforme de Mesures Physiques, and from the French National Research Agency (ANR) under Contract No. ANR-16-CE24-0016-01 and ANR-19-CE47-0010-03. The work done by Probion Analysis was partly supported by the French Research Minister Prorgam: Crédit Impot Recherche.